\documentstyle[12pt]{article}
\newcommand{\beq}{\begin{equation}}
\newcommand{\eeq}{\end{equation}}
\newcommand{\beqa}{\begin{eqnarray}}
\newcommand{\eeqa}{\end{eqnarray}}
\newcommand{\beqan}{\begin{eqnarray*}}
\newcommand{\eeqan}{\end{eqnarray*}}

\newcommand{\ol}{\overline}

\newcommand{\ben}{\begin{enumerate}}
\newcommand{\een}{\end{enumerate}}
\newcommand{\bfl}{\begin{flushleft}}
\newcommand{\efl}{\end{flushleft}}
\newcommand{\ba}{\begin{array}}
\newcommand{\ea}{\end{array}}
\newcommand{\btab}{\begin{tabular}}
\newcommand{\etab}{\end{tabular}}
\newcommand{\bit}{\begin{itemize}}
\newcommand{\eit}{\end{itemize}}

\newcommand{\vs}{\vspace}
\newcommand{\hs}{\hspace}

\def\beq{\begin{equation}}
\def\eeq{\end{equation}}

\begin{document}
\pagestyle{empty}
\thispagestyle{empty}

%\begin{titlepage}
\begin{flushright}
BROWN-HET-1008

BROWN-TA-526

SNUTP 95-102

hep-ph/9510438
\end{flushright}

%\vspace{5mm}
\begin{center}
{\Large \bf   STATUS OF HIGH ENERGY

FOWARD ELASTIC SCATTERING \\}
\vspace{3mm}
{\bf Kyungsik Kang \footnote{Support in part by the USDOE contract DE-FG02-91ER
40688-Task A} and Sung Ku Kim \footnote{Permanent address: Department of
Physics, Ewha Women's University, Seoul 120-750, Korea and
Supported in part by the Korea Science and Engineering Foundation through
Center for Theoretical Physics at Seoul National University, SNU-Brown Exchange
Program and the Ministry of
Education through BSRI-94-2427.}} \\
{\sl Department of Physics, Brown University, Providence, RI 02912, USA \\}
\vspace{3mm}
{\bf   Abstract  \\ }
\end{center}
We present the results of fitting all data for $pp$ and $\bar pp$ scattering
at $\sqrt s \geq 9.7$ GeV and up to the collider energy with various analytic
parametrizations of the elastic forward scattering amplitudes based on the
derivative dispersion relation. It is found that the model containing the
Pomeron and Reggeon terms with 8 parameters has the most preferred
$\chi^2/d.o.f = 1.3$, while the Donnachie and Landshoff model with
5 parameters has $\chi^2/d.o.f = 2.16$ for a data set of 111 experimental
points.
The current data however make no clear preference between the $\ell n s$ and
$\ell n^2 s$ type Pomerons.
%\end{titlepage}
\\

\noindent
{\large \bf I. Complete Survey of Experiments}
\\

It is crucial to select as complete a set of data as possible. No experinmental
group is to be left out, nor any errors to be reduced to account for paucity of
higher
energy results. This type of analysis has been done on different sets of data
depending only on the lowest value of $\sqrt s$ allowed: 9.7 GeV vs. 5
GeV$^{1)}$.
This report deals exclusively with the data set containing 111 experimental
points for the lowest value of $\sqrt s=9.7$ GeV distributed as follows:
58 values of $\sigma_T$, 22 for $\bar pp$ and 36 for $pp$, and 53 values of
$\rho$, 12 for $\bar pp$ and 41 for $pp$. We included the new datas from
UA4.2$^{2)}$, CDF$^{3)}$, and E710$^{4)}$.
Because the majority of precise data is at lower energies
 ($\sqrt s\ < 63$ GeV), we should expect a detailed Regge parametrization.
Also because the newest higher energy experimental data are closer to
standard theoretical expectations, we should expect several theoretical
models to do well.
\\

\noindent
{\large \bf II. Models for the Elastic Forward Scattering Amplitude}
\\

Though we have in principle the exact theory of the strong interactions, QCD,
which can describe, based on the perturbative calculation, the hadron
interactions at short distances, the interacrions at large distance, i.e.,
near forward scattering can not reliably be calculated with the perturbative
QCD. On the other hand, phenomenological models for high energy scattering
based on general principles such as unitarity, analyticity and
crossing-symmetry,
have proven to be successful in understanding or predicting the high-energy
behavior of the hadronic scattering amplitude$^{5)}$. Two such examples are
analytic amplitude models as solutions of the Derivative Dispersion Relation
(DDR)$^{6)}$ derived from the
Sommerfeld-Watson-Regge representation(SWR) and comphensive eikonal
models$^{7)}$, based on QFT expectations or the parton model.

In this report we will concentrate on the analytic amplitude models derived
from DDR.
The crossing-even and odd amplitudes $F_{\pm}$ for $pp$, $\bar pp$ scattering
are defined by

  $$F_{\pm} = \frac{1}{2} (F_{pp} \pm F_{\bar pp})           \eqno(1)$$
where the normalization condition is given by

  $$\sigma_T = \frac{1}{s} Im F(s,t=0)                       \eqno(2)$$
to satisfy the optical theorem.

A Regge pole at $J=\alpha_R(t)$ in the complex $j$-plane gives

  $$F^k_+(s,t) = C^k_+(t)[i-\cot(~\frac{\pi}{2}\alpha^k_+(t)~)]
s^{\alpha^k_+(t)}~~~~~~~~if~~~C~~~is~~~even                    \eqno(3)$$

  $$F^k_-(s,t) = -C^k_-(t)[i+\tan(~\frac{\pi}{2}\alpha^k_-(t)~)]
s^{\alpha^k_-(t)}~~~~~~~~if~~~C~~~is~~~odd                      \eqno(4)$$
If the Regge pole is Exchange-Degenerate, we have

  $$C^k_+ = C^k_-,~~~~~\alpha^k_+ = \alpha^k_-                 \eqno(5)$$

Clearly Regge pole model such as the Donnachie-Landshoff model satisfies
DDRs. A small $\alpha(0)-1=0.08$ is consistent with the slow increase of
$\sigma_T(s)^{8)}$, but would eventually be in conflict with the unitary
condition, i.e., the Froissart bound $\sigma_T(s) \leq C(\ell n s)^2$.
A fully unitary theory must satisfy multiparticle unitary in both
$s$- and $t$-channels. Also the Pomeronchuk theorem, i.e.,
$\sigma^{\bar pp}_T(s)/\sigma^{pp}_T(s) \rightarrow 1$ as $s \rightarrow
\infty$, can be proven rigorously only if total cross-sections increase with
energy$^{9)}$.

If the total cross-section increases with energy, the
leading $j$-plane singularity of $F_+$ is at $j=1$ in the forward direction.
One can then write an analytic parametrization for the contribution of this
asymptotic part from the derivative dispersion relation

  $$F^{p_2}_+(s,o) = i s[A_++B_+(\ell n \frac{s}{s_+}
-i \frac{\pi}{2})^2]                                 \eqno(6)$$
if $\sigma_T(s)$ behaves like $(\ell n s)^2$ at high energies and

  $$F^{p_1}_+(s,o) = i s[A_++B_+(\ell n \frac{s}{s_+}
- i \frac{\pi}{2})]                                  \eqno(7)$$
if $\sigma_T$ increase with energy as $\ell n s$. On the other hand, the
odd-signatured counter part of the Pomeron can also be constructed by that
the difference $\Delta \sigma = \sigma^{\bar pp}_T-\sigma^{pp}_T$ does
not necessarily vanish asymptotically. This implies the leading $j$-plane
singularity of $F_-$ is also at $j=1$ in the forward direction and one
can get the maximal Odderon amplitude from the DDR

  $$F^o_-(s,o) = s[A_-+B_-(\ell n \frac{s}{s_-}-i \frac{\pi}{2})^2]
  ~~~~~~~~~if~~~\Delta \sigma~~ \sim~~ \ell n s
\eqno(8)$$

  $$F^o_-(s,o) = s[A_-+B_-(\ell n \frac{s}{s_-} - i \frac{\pi}{s})]
  ~~~~~~~~~~if~~~\Delta \sigma~~ \rightarrow~~const.           \eqno(9)$$
Equation (8) is the maximal Odderon amplitude if $F_+$ satisfies Eq.(6)
because $\Delta \sigma \leq (\ell n s)^{\beta /2}$ asymptotically if the
cross-section increase as $(\ell n s)^\beta$ with $\beta \leq 2$.

The asymptotic analytic amplitude model can then be constructed in various
form of

  $$F_+ = F^{p_i}_+(s,o) + \sum_k F^{(k)}_+(s,o)          \eqno(10)$$
  $$F_- = F^o_-(s,o) + \sum_k F^{(k)}_-(s,o)              \eqno(11)$$
where $F^{(k)}_{\pm}$ represents the Regge amplitudes.
\\

\noindent
{\large \bf III. Comparison with Experiment}
\\

We now present the results of our comprehensive $\chi^2$-fits to 111 high
energy data points by theoretical models with and without the
Odderon terms along with the predictions at the LHC energy.

Depeding on the choices of various terms from the expressions (10) and (11), we
studied the following 10 models:
\\
\noindent
(a) Model A1: The Block-Kang-White model$^{1)}$, $P_1+RND_++RND_-$, where $P_1$
is the $\ell n s$-type Pomeron term and $RND_{\pm}$ represent non-degenerate
Regge terms. Since $A_+-B_+\ell n s_+$ is fixed, this model has 6 (=3+2+2-1)
free parameters.
\\
\noindent
(b) Model A2: Modification of Model A1 by replacing $P_1$ by the $(\ell n
s)^2$-type Pomeron term, $P_2+RND_++RND_-$ with 7 parameters.
\\
\noindent
(c) Model A3: $P_2+RD$ with 5 parameters, where $RD$ is the exchange degenerate
Regge term.
\\
\noindent
(d) Model B1: $P1+RD+RND_++RND_-$ with 8 free parameters.
\\
\noindent
(e) Model B2: Modification of Model B1 by replacing $P_1$ by $P_2$, i.e.,
$P_2+RD+RND_++RND_-$ with 9 free parameters.
\\
\noindent
(f) Model E1: The maximal Odderon model,
$P_2+O+RD+RND_++RND_-$ with 12 free parameters.
\\
\noindent
(g) Model E2: The maximal Odderon
model with one exchange degenerate Regge term, $P_2+O+RD$ with 8 parameters.
\\
\noindent
(h) Model E3: The maximal Odderon model without the exchange degenerate Regge
term,
 $P_2+O+RND_++RND_-$ with 10 parameters.
\\
\noindent
(i) Model F1: The model with the bare
Pomeron plus the nondegenerate Regge terms,

\hs{20mm} $C_1 \hs{2mm} s^{A_1} \hs{2mm} +
\hs{2mm} C_2 \hs{2mm} s^{- A_2} \hs{2mm} +
\hs{2mm} C_3 \hs{2mm} s^{- A_3}$
\hs{5mm} for \hs{3mm} $\sigma_{p\ol{p}}$,

\hs{20mm} $C_1 \hs{2mm} s^{A_1} \hs{2mm} +
\hs{2mm} C_2 \hs{2mm} s^{- A_2} \hs{2mm} -
\hs{2mm} C_3 \hs{2mm} s^{- A_3}$
\hs{5mm} for \hs{3mm} $\sigma_{pp}$.
\\
\noindent
This model has 6 parameters.
\\
\noindent
(j) Model F2: The Donnachie-Landshoff (DL) Model$^{8)}$, which is a
modification of Model F1 by setting $A_2=A_3$. This model is similar to Model
A3 and
has 5 free parameters.

The results of the fits for the parameters are given below in Table along with
the predictions of $\rho$ and $\sigma_T$ at LHC ($\sqrt s$=14 Tev).
Fig. 1 shows fits to Model B1 and Fig. 2 exhibits the fits to Model F2, the
Donnachie-Landshoff Regge model.
\\

\noindent
{\large \bf IV. Conclusions }
\\
\noindent
(a) Model B1, $P_1 + RD + RND$, and Model B2, $P_2 + RD + RND$,
have the most prefered fits with $\chi^2/d.o.f = 1.3$
to the data for $\sqrt s \geq 9.7$ GeV. There are, however, no significant
differences between the two models: the current data are consistent with either
$\ell n s$ or $\ell n^2 s$ increase of $\sigma_T$.
\\
\noindent
(b) All of the models considered above based on the asymptotic analytic
amplitudes reproduce the data more or less equally well, i.e., $\chi^2/d.o.f
\simeq 1.3 \sim 1.45$, considering the quality of the data from different
experiments. In particular, the maximal Odderon models E1 and E2 give
comparable fits as
those of Models B1 and B2 so that there seems to be little support for the
Odderon in the current data.
\\
\noindent
(c) The Regge model F2 has relatively high $\chi^2/d.o.f$ but the Model F1,
which is a modified version of F2 by relaxing the exchange degeneray
assumption, has significantly improved and comparable $\chi^2/d.o.f$ value
as those of Model A1 or A2.
\\
\noindent
(d) For the asymptotic behavior of $\sigma_T(s)$ extracted from this study and
a
comprasion with the phenomenological fits made in the 1994 Review of Particle
 Properties$^{10)}$, see Kang's talk$^{11)}$.
\\
\noindent
(e) We denote that the cofficient $C$'s of exchange degenerate Regge terms
have values with huge errors in Model B1 and B2: 6042.5$\pm$2261.6 and
8857.2$\pm$7424.6, respectively. Thus we should need more low energy data by
lowering the cut-off below 9.7 GeV, if their errors are to be reduced and
improved.
\\
\noindent

One of us (SKK) would like to thank Brown University for warm
hospitality extended to him during his sabbatical stay during the
academic year 1994-95.
\\

\noindent
{\large \bf V. References }
\\
\begin{description}
\item[{\rm [1]}] K. Kang, Nucl. Phys. {\bf B12} (Proc. Suppl. 1990) 64;
M. M. Block, K. Kang and A. R. White,
Mod. Phys. {\bf A7} (1992) 4449;
K. Kang, P. Valin and A. R. White,
in Proceedings of the International Conference (V Blois Workshop) on Elastic
and Diffractive Scattering, 8-12 June 1993, Brown university;
K. Kang, P. Valin and A. R. White,
Nouco Cimento {\bf 107A} (1994) 2103.
\item[{\rm [2]}] UA4/2 Collab, C. Augier et al.,
Phys. Lett. {\bf B316} (1993) 448: CERN/PPE 96-160, 6 Oct. 1994.
\item[{\rm [3]}] CDF Collab, F. Abe et al., Phys. Rev. {\bf D50} (1994) 5550.
\item[{\rm [4]}] S. Sadr, in Proc. of Vth Blois Workshop.
\item[{\rm [5]}] K. Kang and B. Nicolescu, Phys. Rev. {\bf D11} (1975) 2461;
M. M. Block and R. N. Cahn, Rev. Mod. Phys. {\bf 57} (1985) 563.
\item[{\rm [6]}] V. N. Gribov and A. A. Migdal, Sov. F. Nucl. Phys. {\bf 8}
(1969) 583; K. Kang and B. Nicolescu (Ref. 5).
\item[{\rm [7]}] H. Cheng, J. K. Walker and T. T. Wu, Phys. Lett. {\bf 44B},
(1973) 97; C. Bourreley,
J. Soffer and T. T. Wu, Phys. Rev. {\bf D19} (1979) 3249; P. L'Heureux,
B.Margolis and P. Valin, Phys. Rev. {\bf D32} (1985) 1681.
\item[{\rm [8]}] P. D. B. Collins, F. D. Gault and A. Martin, Nucl. Phys. {\bf
B85} (1977) 141;
A. Donnachie and P. V. Landshoff, Nucl. Phys. {\bf B267} (1986) 657; Phys.
Lett. {\bf B296} (1992) 227.
\item[{\rm [9]}] T. Kinoshita, Perspectives in Mod. Phys., Ed. R. Marshak (John
Wiley and Sons, 1966); R. J. Eden, Phys. Rev. Lett. {\bf 16} (1966) 39;
G. Grunberg and T. N. Truong, Phys. Rev. Lett. {\bf 31} (1973) 63.
\item[{\rm [10]}] Particle Data Group, K. Hikasa et al., Phys. Rev. {\bf D50}
(1994) 1173.
\item[{\rm [11]}] K. Kang, these Proceedings.
\end{description}

\noindent
{\large \bf Figure Captions}
\\
\begin{description}
\item[Fig. 1]  Model (B1) fits to (a) $\sigma_T$ and (b) $\rho$
\item[Fig. 2]  Model (F2) fits to (a) $\sigma_T$ and (b) $\rho$
\end{description}

\newpage
\footnotesize
\begin{table}
\begin{center}
\begin{tabular}{|c|c|c|c|c|c|}
\hline
Modles
& A1
& A2 & A3 & F1 &F2\\
\hline
%Parameters & $P_1+RND$
%& $P_2+RND$ & $P_2+RD$ & &$A_2=A_3$ \\
\hline
$A_{+}$ &21.241 & 28.286 &39.062 & & \\
\hline
$B_{+}$ &6.8241 & 0.22793 &0.36909& & \\
\hline
$\ln \sqrt{s_{+}}$& 3.9465 & 0.19882 & 2.2001& & \\
\hline
$A_{-}$ & 0. & 0. & 0. & &  \\
\hline
$B_{-}$ & 0. & 0. & 0. & &  \\
\hline
$\ln \sqrt{s_{-}}$ & 0. & 0. & 0. & &  \\
\hline
$C_{+}$ &104.93 & 51.914 & 0. &$C_1:18.952$ & 22.782 \\
\hline
$\alpha_{+}$ &0.80213 &0.62470 &  &$A_1:0.093272$&0.076403 \\
\hline
$C_{-}$ &34.498&35.544&0&$C_2:62.182$&$C_2+C_3:98.278$ \\
\hline
$\alpha_{-}$ &0.45051 &0.44443 &  &$A_2:0.35989$ &0.47349 \\
\hline
$C$ & 0. & 0. & 37.387 &$C_3:35.677$&$C_2-C_3:52.055$ \\
\hline
$\alpha$ &  &  &0.43500&$A_3:0.55633$& \\
\hline
$\chi^2$ &146.0307&142.4876&154.1655&142.8956&229.2668 \\
\hline
$\chi^2/d.o.f $ & 1.39077 & 1.37007 &1.45439 & 1.36091 & 2.16289 \\
\hline
$\rho_{p\ol{p}}$ (546) & 0.1283 & 0.1361 & 0.1507 & 0.1395 & 0.1182 \\
\hline
. \hs{5mm}(1800) & 0.1197 & 0.1351 & 0.1543 & 0.1448 & 0.1199 \\
\hline
. \hs{4mm}(14000) &0.0994 & 0.1244 & 0.1446 & 0.1471 & 0.1205 \\
\hline
$\rho_{pp}$ (546) &0.1275 & 0.1354 & 0.1498 & 0.1388 & 0.1163 \\
\hline
. \hs{5mm}(1800) & 0.1196 & 0.1350 & 0.1542 & 0.1446 & 0.1194 \\
\hline
. \hs{4mm}(14000) & 0.0994 & 0.1244 & 0.1446 & 0.1471 & 0.1205 \\
\hline
$\sigma_{p\ol{p}}$ (546) & 62.10 & 62.18 & 63.06 & 62.11 &59.94 \\
\hline
. \hs{5mm}(1800) & 75.09 &  76.46 & 79.57 & 77.01 & 71.70 \\
\hline
. \hs{4mm}(14000) &100.1 & 107.4 & 117.8 & 112.5 & 98.00 \\
\hline
$\sigma_{pp}$ (546) &62.03 &  62.12 & 63.00 & 62.05 & 59.82 \\
\hline
. \hs{5mm}(1800) &75.07 & 76.44 & 79.55 & 76.99 & 71.66 \\
\hline
. \hs{4mm}(14000) &100.1&107.4&117.8&112.5&97.99 \\
\hline
\hline
Modles
& E1
& E2 & E3 & B1&B2 \\
\hline
%Parameters &  &
%&  & &  \\
\hline
$A_{+}$ & 28.830 & 39.269 & 29.330 &6.4816 &28.738 \\
\hline
$B_{+}$ & 0.23280 & 0.36182 & 0.23010 &8.2255&0.24247 \\
\hline
$\ln \sqrt{s_{+}}$ & 0.26108 & 2.1936 & 0.29304 &4.1055 &0.43383 \\
\hline
$A_{-}$ & - 0.47475 & 12.002 & 0.28456 &0. &0. \\
\hline
$B_{-}$ & - 0.12010 & - 0.00040921 & - 0.036942 &0. &0. \\
\hline
$\ln \sqrt{s_{-}}$ & 6.1277 & 89.530 & 6.0487 &0. &0. \\
\hline
$C_{+}$ & 157.28 & 0. & 54.878 &126.05 &42.091 \\
\hline
$\alpha_{+}$ &0.58499 &  &0.59521 &0.85242 &0.66971 \\
\hline
$C_{-}$ & 164.03 & 0. & 112.94 &25.689 &25.249 \\
\hline
$\alpha_{-}$ & 0.47052 &  & 0.11584 &0.49582 &0.49882 \\
\hline
$C$ & - 103.03 & 44.817 & 0. &6042.5 &8857.2 \\
\hline
$\alpha$ & 0.57285  & 0.37601 & &-1.1661 &-1.2384 \\
\hline
$\chi^2$ & 133.2673 & 148.3081 & 134.1526 &133.9024 &134.1805 \\
\hline
$\chi^2/d.o.f $ & 1.34613 & 1.43988 & 1.32824 &1.30002 &1.31550 \\
\hline
$\rho_{p\ol{p}}$ (546) & 0.1343 & 0.1374 & 0.1297 &0.1337 &0.1384 \\
\hline
. \hs{5mm}(1800) & 0.1499 & 0.1395 & 0.1344 &0.1272 &0.1380 \\
\hline
. \hs{4mm}(14000) & 0.1824 & 0.1300 & 0.1385 &0.1076 & 0.1273 \\
\hline
$\rho_{pp}$ (546) & 0.1394 & 0.1592 & 0.1412 &0.1325 &0.1371 \\
\hline
. \hs{5mm}(1800) & 0.1222 & 0.1653 & 0.1349 &0.1269 &0.1377 \\
\hline
. \hs{4mm}(14000) & 0.07015 & 0.1567 & 0.1100 &0.1076 &0.1272 \\
\hline
$\sigma_{p\ol{p}}$ (546) & 62.21 & 63.06 & 62.28 &62.32 &62.25 \\
\hline
. \hs{5mm}(1800) & 75.99 & 79.28 & 76.30 &76.09 &76.82 \\
\hline
. \hs{4mm}(14000) & 106.0 & 116.8 & 106.8 &103.6 &108.8 \\
\hline
$\sigma_{pp}$ (546) & 63.01 & 62.60 & 62.39 &62.23 &62.16 \\
\hline
. \hs{5mm}(1800) & 78.28 & 78.85 & 76.97 &76.06 &76.79 \\
\hline
. \hs{4mm}(14000) & 111.2 & 116.4 & 108.4 &103.6 &108.8 \\
\hline
\end{tabular}

\vs{3mm}
{\large \bf Table} \\
\end{center}
\end{table}

\end{document}